\documentclass[10pt]{article}
\usepackage{amsmath,amsfonts,xypic}
\usepackage{amssymb}
\usepackage{amscd}
\usepackage[all]{xy}
\advance \textheight by \topskip
\makeatletter
\@addtoreset{equation}{section}
 
\makeatother

\newtheorem{theorem}{Theorem}[section]

\newtheorem{definition}[theorem]{Definition}
\newtheorem{example}[theorem]{Example}

\newtheorem{proposition}[theorem]{Proposition}

\newcommand{\beq}{\begin{equation}}
\newcommand{\eeq}{\end{equation}}
\newcommand{\beqa}{\begin{eqnarray}}
\newcommand{\eeqa}{\end{eqnarray}}
\newcommand{\noi}{\noindent}

\newcommand{\g}{{\mathfrak g}}

\newcommand{\totimes}{ \ \hat  \otimes \ }

\def\>{\rangle}
\def\<{\langle}

\begin{document}
\title{Ternary algebras and groups}

\author{
{\sf  M. Rausch de Traubenberg }\thanks{e-mail:
rausch@lpt1.u-strasbg.fr} \\
{\it
Laboratoire de Physique Th\'eorique, CNRS UMR  7085,
Universit\'e Louis Pasteur}\\
{\small {\it  3 rue de
l'Universit\'e, 67084 Strasbourg Cedex, France}}}
\date{\today}
\maketitle

\begin{abstract}
We construct  explicitly    groups
associated to 
specific ternary algebras which extend the Lie (super)algebras 
(called
Lie algebras of order three). It turns out that the natural
variables which appear in this construction are variables which
generate the three-exterior algebra.  
An explicit matrix representation of a group associated to a peculiar
Lie algebra of order three is constructed considering matrices with entry
which belong to the three exterior algebra. 
\end{abstract}

\section{Introduction}
Most of the laws of physics are based on algebras {\it i.e.} on mathematical
structures equipped with a  binary product. That is, if one considers an 
algebra ${\cal A}$ with a product $m_2$ we have  for any 
$a\otimes b \in  {\cal A} \otimes {\cal A},
m_2(a\otimes b) \in {\cal A}$. Among algebras the Lie algebras and the Lie
superalgebras are of special 
importance. The former  allow a description of the space-time
and internal symmetries in particle physics although the latter lead  to 
a supersymmetric extension of symmetries. In both cases the Lie
(super)algebras correspond to the symmetries of a physical system at
the infinitesimal level.  However, in these two cases
one  is able to define symmetries at
a finite level. The corresponding symmetries are associated to Lie
groups or Lie supergroups.

The ternary algebras have been considered in physics only occasionally
(see for instance \cite{bg1,bg2,k1,k2,k3,r} and references therein).
For some mathematical references one can see 
\cite{f,g,mv}.  A ternary 
algebra ${\cal A}$ is an algebraic structure equipped with a ternary
product $m_3: {\cal A} \otimes {\cal A} \otimes {\cal A} \to {\cal A}$.

In \cite{rs1,rs2} an   $F-$arry algebra which can be seen as a possible
generalisation of Lie (super)algebras have been considered and named
Lie algebra of order $F$. 
A Lie algebra of order $F$  admits a $\mathbb Z_F-$grading
($F=3$ in this paper), the zero-graded part
being a Lie algebra. An $F-$fold symmetric product (playing the role of
the anticommutator in the case $F=2$)  expresses the zero graded  part
in terms of the non-zero graded part. This means that part of the algebra
is a (binary)  algebra and part of the algebra is a ternary algebra (when
$F=3$). 
 Subsequently, a  specific Lie algebra or order $3$,
leading to a non-trivial extension of the Poincar\'e algebra,
  has been studied 
together with its implementation in Quantum Field Theory  
\cite{cubic1,cubic2,pform}. 
A general study of the possible non-trivial extensions of the
Poincar\'e algebra in $(1+3)-$dimensions 
has been undertaken in \cite{grt1,grt2}.
However, all these mathematical structures have been considered at
the level of  algebras {\it i.e.} at the level of infinitesimal
transformations and no groups associated to Lie algebras of order $3$
were considered. At a first glance these  two structures seem to
be incompatible since for a Lie algebra of order 3 for some elements
only the product of three elements is defined although for a group
the product of two elements is always defined. 

In this paper we show that
in fact a specific ternary group that we might  call a Lie
group of order $3$ can be associated to a Lie algebra of order $3$.
We show that one possible way to associate
a group to a given Lie algebra of order 3 is to endow the
universal enveloping algebra of a Lie algebra of order 3 with
a Hopf algebra structure. Then we identify the parameters of the
transformation. It turns out that these parameters are strongly related
to the $3-$exterior algebra. The $3-$exterior algebra being
the algebra generated by canonical generators $\theta^i$ satisfying
\cite{roby}

$$
 \theta^i \theta^j \theta^k + 
\theta^j \theta^k \theta^i + 
\theta^k \theta^i \theta^j +
  \theta^i \theta^k \theta^j + 
\theta^j \theta^i \theta^k + 
\theta^k \theta^j \theta^i=0.
$$

The content of this paper is the following. 
In section 2 we recall the definition of  Lie algebras of order 3 and 
give some examples. We then  define its universal enveloping algebra
and  endow it  with a Hopf algebra structure.
Then the consideration of the Hopf dual   enable us to define 
the parameters of the transformations which 
turn out to be related to  the $3-$exterior algebra \cite{roby}.
In section 3 we construct explicitly a group associated to 
a specific Lie algebra of order 3 by considering matrices
with entries which belong to the three exterior algebra.

\section{Lie algebras of order 3}
\subsection{Definition and examples}
The general definition of  Lie algebras of order $F$, was given in 
\cite{rs1,rs2}
together with an inductive way to construct Lie algebras of order $F$ 
associated 
with {\it any} Lie algebra or Lie superalgebra.  We recall here the main
 results 
useful for the sequel.
Let  $F$  be a positive integer and define 
$q=e^{\frac{ 2 \pi i}{F}}$.  We consider $\g$ a complex vector space and 
$\varepsilon $ an automorphism
of $\g$ satisfying $\varepsilon^F=1$. 
Set $\g_i \subseteq \g$ the eigenspace corresponding to
the eigenvalue $q^i$ of $\varepsilon$. 
Then, we have $\g=\g_0 \oplus \cdots  \oplus \g_{F-1}$.\\

\begin{definition}
\label{def-flie}
Let $F\in\mathbb{N}^*$.
A ${\mathbb Z}_F$-graded ${\mathbb C}-$vector space  ${\mathfrak{g}}= {\mathfrak{g}}_0 \oplus
{\mathfrak{g}}_1\oplus
{\mathfrak{g}}_2 \dots \oplus
{\mathfrak{g}}_{F-1}$ 
is called a complex Lie algebra of order $F$ if
\begin{enumerate}
\item $\mathfrak{g}_0$ is a complex Lie algebra.
\item For all $i= 1, \dots, F-1 $, $\mathfrak{g}_i$ is a representation of $\mathfrak{g}_0$. If $X \in \g_0, Y \in \g_i$ then $[X,Y]$ denotes
the action of $X \in \g_0$ on
$Y \in \g_i$ for all $i=1,\cdots F-1$.
\item  For all $i=1,\dots,F-1$ there exists  an $F-$linear,  $\mathfrak{g}_0-$equivariant  map
 $$\{ \cdots \} : {\cal S}^F\left(\mathfrak{g}_i\right)
\rightarrow \mathfrak{g}_0,$$
 where  ${ \cal S}^F(\mathfrak{g}_i)$ denotes
the $F-$fold symmetric product of $\mathfrak{g}_i$, satisfying the following (Jacobi) identity
\beqa
\label{eq:J}
&&\sum\limits_{j=1}^{F+1} \left[ Y_j,\left\{ Y_1,\dots,
Y_{j-1},
Y_{j+1},\dots,Y_{F+1}\right\} \right] =0, \nonumber
\eeqa
\noi
for all $Y_j \in \mathfrak{g}_i$, $j=1,..,F+1$. 
\end{enumerate}
\end{definition}

\medskip

By definition when $F=1$,  $\g=\g_0$ is a Lie algebra and when
$F=2$, $\g=\g_0  \oplus \g_1$ is a Lie superalgebra. Thus Lie
algebra of order three can be seen as some possible generalisation of Lie
(super)algebras. It can be noticed that for any $k=1,\ldots,F-1$, the 
$\mathbb Z_F-$graded vector spaces ${\mathfrak{g}}_0\oplus{\mathfrak{g}}_k$ 
satisfy all the properties in Definition \ref{def-flie} and thus
are  Lie algebras of order $F$.  We  call these types of algebras 
\textit{elementary Lie algebras of order $F$}. 

\begin{definition}\label{representation}
A representation of an elementary Lie algebra of order $F$
is  a linear map
$\rho : ~ \g=\g_0 \oplus \g_1 \to \mathrm{End}(V)$, 
such that  (for all $X_i \in \g_0, Y_j \in \g_1$)

\beqa      
\label{eq:rep}
\begin{array}{ll}
& \rho\left(\left[X_1,X_2\right]\right)= \rho(X_1) \rho(X_2)- 
\rho(X_2)\rho(X_1) \cr
& \rho\left(\left[X_1,Y_2\right]\right)= \rho(X_1) \rho(Y_2)- 
\rho(Y_2)\rho(X_1) \cr
& \rho\left( \left\{Y_1.\cdots,Y_F\right\}\right)=
 \sum \limits_{\sigma \in S_F}
\rho\left(Y_{\sigma(1)}\right) \cdots \rho\left(Y_{\sigma(F)}\right) \cr
\end{array}
\eeqa

\noindent
($S_F$ being the group of permutations of $F$ elements).
\end{definition}

\noi
By construction the vector space $V$ is  $\mathbb Z_F-$graded
 $V= V_0 \oplus \cdots \oplus V_{F-1}$ 
 and for all $a \in\{0,\cdots, F-1\}$, $V_a$ is a $\g_0-$module and we have 
$\rho(\g_1) (V_a) \subseteq V_{a+1}$.
\medskip

From now on we only consider elementary Lie algebras of order $3$.
It can be observed that part of the algebra is binary and part of the
algebra is ternary.  Indeed we have by Definition \ref{def-flie} the
following products: $m_1: \g_0 \otimes \g_0 \to \g_0, \ 
m_2: \g_0 \otimes \g_1 \to \g_1, \ m_3: \g_1 \otimes \g_1 \otimes \g_1
\to \g_0$. Furthermore,
there is essentially two types of Lie algebras
of order 3. The first types are the  Lie algebras of order 3 related
to Lie algebras of order 3  obtained from a Lie (super)algebra 
through the inductive 
 process \cite{rs2}. One example of this types of algebras is
 the cubic extension of the Poincar\'e algebra

 \begin{example}
\label{FP}
Let $\g_0 = \left< L_{\mu \nu }, P_\mu\right>$ be the  Poincar\'e
 algebra in $D-$dimensions and  
$\g_1=\left<V_\mu \right>$ be the $D-$dimensional vector representation of $\g_0$.
The brackets
\beqa\label{fp-bracket}
 \left[L_{\mu \nu }, L_{\rho \sigma}\right]&=&
\eta_{\nu \sigma } L_{\rho \mu }-\eta_{\mu \sigma} L_{\rho \nu} + \eta_{\nu \rho}L_{\mu \sigma}
-\eta_{\mu \rho} L_{\nu \sigma},\nonumber \\
\left[L_{\mu \nu }, P_\rho \right]&=& \eta_{\nu \rho } P_\mu -\eta_{\mu \rho } P_\nu,  \ 
\left[L_{\mu \nu }, V_\rho \right]= \eta_{\nu \rho } V_\mu -\eta_{\mu \rho } V_\nu, \
\left[P_{\mu}, V_\nu \right]= 0, \nonumber \\ 
\{ V_\mu, V_\nu, V_\rho \}&=&
\eta_{\mu \nu } P_\rho +  \eta_{\mu \rho } P_\nu + \eta_{\rho \nu } P_\mu,
\nonumber
\eeqa
\noi
with the metric $\eta_{\mu \nu}=\rm{diag}(1,-1,\cdots,-1)$, endow 
$\g=\g_0 \oplus \g_1$ with 
an elementary Lie algebra of order $3$ structure.
\end{example}

\noi
The second types of Lie algebras of order 3 are obtained as a sub-algebra
of Lie algebras of order 3 defined by some matrix representation.

\begin{example} \label{mat} 
Let 
$\mathfrak{gl}(m_1,m_2,m_3)$ and 
$\mathfrak{gl}_{\mathrm{el}}(m_1,m_2,m_3)$
be the set of  $(m_1 + m_2 +m_3)
\times (m_1 + m_2 + m_3)$ matrices of the form

\beqa\label{gl}
\begin{array}{ll}
\mathfrak{gl}_{\mathrm{el}}(m_1,m_2,m_3) = \left\{
\begin{pmatrix} a_0&b_1&0\\
                  0&a_1&b_2\\
                  b_0&0&a_2
  \end{pmatrix} \right\},&
\mathfrak{gl}(m_1,m_2,m_3)=\left\{
\begin{pmatrix} a_0&b_1&c_2\\
                  c_0&a_1&b_2\\
                  b_0&c_1&a_2
  \end{pmatrix}\right\},
\end{array}
\eeqa
\noi
with $a_0 \in \mathfrak{gl}(m_1), a_1 \in \mathfrak{gl}(m_2),a_2 \in 
\mathfrak{gl}(m_3),$  $
b_1 \in {\cal M}_{m_1,m_2}(\mathbb C), b_2\in  {\cal M}_{m_2,m_3}(\mathbb C),
 b_0 \in
{\cal M}_{m_3,m_1}(\mathbb C)$,
and   $
c_0 \in {\cal M}_{m_2,m_1}(\mathbb C), c_1\in  {\cal M}_{m_3,m_2}(\mathbb C),
 c_2 \in
{\cal M}_{m_1,m_3}(\mathbb C)$.
A basis of these sets of matrices
 can be constructed as follow. Consider the 
$ (m_1 + m_2 + m_3)^2$ canonical  matrices $e_{I}{}^{J}, 
\ 1 \le I, J \le m_1 +m_2
+m_3$. With the following convention for the indices 
$ 1 \le i,j \le m_1, \  m_1 + 1 \le i',j' \le m_1+m_2,\ 
 m_1 + m_2 + 1 \le i'',j'' \le m_1+m_2 + m_3$,
the generators  are given by

$$\begin{array}{lll}
e_{i}{}^{j}\text{ for } \mathfrak{gl}(m_1),&
e_{i'}{}^{j'}\text{ for } \mathfrak{gl}(m_2),&
e_{i''}{}^{j''}\text{ for } \mathfrak{gl}(m_3),\\
e_{i}{}^{j'} \text{ for } {\cal M}_{m_1,m_2}(\mathbb C),&
e_{i'}{}^{j''} \text{ for } {\cal M}_{m_2,m_3}(\mathbb C),&
e_{i''}{}^{j} \text{ for } {\cal M}_{m_3,m_1}(\mathbb C),\\
e_{i'}{}^{j} \text{ for } {\cal M}_{m_2,m_1}(\mathbb C),&
e_{i''}{}^{j'} \text{ for } {\cal M}_{m_3,m_2}(\mathbb C),&
e_{i}{}^{j''} \text{ for } {\cal M}_{m_1,m_3}(\mathbb C).\\

\end{array}$$

\noi
Writing $\mathfrak{gl}(m_1,m_2,m_3) = 
\mathfrak{gl}(m_1,m_2,m_3)_0 \oplus \mathfrak{gl}(m_1,m_2,m_3)_1
\oplus  \mathfrak{gl} (m_1,m_2,m_3)_2$ and  
$\mathfrak{gl}_{\mathrm{el}}(m_1,m_2,m_3) = 
\mathfrak{gl}_{\mathrm{el}}(m_1,m_2,m_3)_0 
\oplus \mathfrak{gl} _{\mathrm{el}}(m_1,m_2,m_3)_1$ 
we denote generically by $X_I{}^J$ the generators of grade zero  
$Y_I{}^J$ the generators of grade one and $Z_I{}^J$ those of grade two, and 
the brackets read 

\beqa \label{gl3}
[X_{I}{}^{J}, X_{K}{}^{L}]&=& \delta^J{}_K X_I{}^L-\delta^L{}_I X_K{}^J,
\nonumber \\
\left[X_{I}{}^{J}, Y_{K}{}^{L}\right]&=& \delta^J{}_K Y_I{}^L
-\delta^L{}_I Y_K{}^J,  \nonumber \\
\left[X_{I}{}^{J}, Z_{K}{}^{L}\right]&=& \delta^J{}_K Z_I{}^L
-\delta^L{}_I Z_K{}^J,  \nonumber \\
\left\{Y_I{}^J,Y_K{}^L,Y_M{}^N\right\}&=&
\delta^J{}_K \delta^L{}_M X_I{}^N +
\delta^N{}_I \delta^J{}_K X_M{}^L +
\delta^L{}_M \delta^N{}_I X_K{}^J  \\
&+& \delta^J{}_M \delta^N{}_K X_I{}^L +
\delta^N{}_K \delta^L{}_I X_M{}^J +
\delta^L{}_I \delta^J{}_M X_K{}^N,  
\nonumber \\
\left\{Z_I{}^J,Z_K{}^L,Z_M{}^N\right\}&=&
\delta^J{}_K \delta^L{}_M X_I{}^N +
\delta^N{}_I \delta^J{}_K X_M{}^L +
\delta^L{}_M \delta^N{}_I X_K{}^J \nonumber \\
&+& \delta^J{}_M \delta^N{}_K X_I{}^L +
\delta^N{}_K \delta^L{}_I X_M{}^J +
\delta^L{}_I \delta^J{}_M X_K{}^N,  
\nonumber 
\eeqa
this shows that $\mathfrak{gl}(m_1,m_2,m_3)$ 
(resp. $\mathfrak{gl}_{\mathrm{el}}(m_1,m_2,m_3)$) is endowed with the structure
of Lie algebra of order three (resp. the structure of
 elementary Lie algebra of order three). 
\end{example}

The algebra of Example 
\ref{FP} was firstly introduced in \cite{rs1} and
its implementation in Quantum Field Theories was realised in $4-$dimensions
in \cite{cubic1,cubic2} and in $D-$dimensions in relations to generalised
gauge fields or $p-$forms in \cite{pform}. 
A general classification of Lie algebra of order $3$ extending non-trivially
the Poincar\'e algebra in $4-$dimensions was undertaken in \cite{grt1,grt2}.
However, even if invariant
Lagrangian were constructed all was done at the level of algebras
{\it i.e.} of infinitesimal transformations and nothing were said
at the level of groups.  Indeed,  by definition, a Lie
algebra of order 3 is partially a ternary algebra since
the product of three elements of $\g_1$ is defined and not
the product of two. This seem to be {\it a priori} in contradiction
with groups since for a group a product of two elements  is always defined.
We will see in the next subsection how one can evade this contradiction
and construct a group associated to the ternary algebras we are considering.

\subsection{Hopf algebras associated to elementaries Lie algebras of order 3}
\label{hopf-sect}
In this subsection, we show that the introduction
of Hopf algebras is a possible way to
construct a group associated to a given Lie algebra of order 3. We
also show that the parameters of the transformation associated to 
an element of $\g_1$ belong to the $3-$exterior algebra (see the appendix
for definition).

Let $\g = \g_0 \oplus \g_1$ be a given Lie algebra of order 3.
We denote generically by $X$ the elements of $\g_0$ and
by $Y$ the elements of $\g_1$.
To endow $\g$ with a Hopf algebra structure we need first to
define ${\cal U}(\g)$ the  enveloping algebra associated to $\g$.
Consider the tensor algebra

$$T(\g) = \mathbb C \oplus \g \oplus (\g \otimes \g) \oplus \cdots \ .$$

\noi
This is an associative algebra with multiplication given
for any $t_1, t_2 \in T(\g)$ 
 by 
$t_1 \otimes t_2 \in T(\g)$. Now, in order to assume
the multiplication given in Definition \ref{def-flie} we
define (for any $X_1, X_2 \in \g_0, Y_1, Y_2, Y_3 \in \g_1$)
the two-sided ideal ${\cal I}$   generated by

\beqa
&&X_1\otimes X_2 - X_2 \otimes  X_1 -[X_1,X_2], \nonumber \\
&&X_1\otimes Y_2 - Y_2 \otimes  X_1 -[X_1,Y_2], \nonumber \\
&&Y_1 \otimes Y_2 \otimes Y_3 + 
Y_2 \otimes Y_3 \otimes Y_1 + 
Y_3 \otimes Y_1 \otimes Y_2 + \nonumber \\
&&Y_1 \otimes Y_3 \otimes Y_2 + 
Y_2 \otimes Y_1 \otimes Y_3 + 
Y_3 \otimes Y_2 \otimes Y_1 - \left\{Y_1,Y_2,Y_2\right\}. \nonumber 
\eeqa 

\noi
The enveloping algebra of $\g$ (in fact it is universal but this will
be shown elsewhere \cite{gr}) is then defined by 
${\cal U}(\g) = T(\g)/{\cal I}$. This means in particular that
in the associative algebra ${\cal U}(\g)$ the relation
given in Definition \ref{def-flie} are satisfied.
Furthermore, as in the Lie (super)algebra cases, a basis of
${\cal U}(\g)$ can be given. This is an analogous of the  
the Poincar\'e-Birkhoff-Witt theorem in this case. Indeed,
it can be proven \cite{gr} that
we have the following isomorphism of vector space

\beqa
\label{pbw}
{\cal U} (\g) \cong S(\g_0) \otimes \Lambda_3(\g_1)
\eeqa

\noi
where  $S(\g_0)$ is the symmetric algebra over $\g_0$ and
$\Lambda_3(\g_1)$ is the three-exterior algebra on $\g_1$
(see the appendix for definition).
Since the composition of the natural map 
$\g  \to T(\g)$ with the canonical projection
$T(\g)\to  {\cal U}(\g)$ gives $\g \subset  {\cal U}(\g)$  we identify
$\g$ with its image under this map. Thus, if we denote 
$\{ X_a, 1 \le a \le \text{dim } \g_0=n_0\}$ (resp. 
$\{ Y_i, 1 \le i \le \text{dim } \g_i=n_1\}$ 
a basis of $\g_0$ (resp. $\g_1$) a basis of ${\cal U}(\g)$ is given by 
the elements of the form \cite{gr}

\beqa
\label{basis}
g_{\vec a, I_\ell}=
\frac{X_1^{a_1}}{a_1 !} \cdots \frac{ X_{n_0}^{a_{n_0}}}{a_{n_0}!}
Y_{I_\ell}, \ \ (a_1, \cdots, a_{n_0}) \in \mathbb N^{n_0}, \ \ 
\ell \in \mathbb N, \ \ 
I _\ell \in \left\{1,\cdots,n_1\right\}^{\ell } \setminus I_{\ell,n_1}.
\eeqa

\noi
where $Y_{I_\ell}= Y_{(i_1,\cdots,i_\ell)}= Y_{i_1} \cdots Y_{i_\ell}$.
(See the appendix for the definition of $I_{\ell,n_1}$.)
An element $Y_{I_\ell}$ with $I _\ell \in \left\{1,\cdots,n_1\right\}^{\ell } \setminus I_{\ell,n_1}$
is called a Roby element.

As in the usual Lie (super)algebra cases, ${\cal U}(\g)$ can be
endowed with a Hopf algebra structure. Although the algebra structure
depends on the brackets defining the Lie algebra of order $3$ we
are considering, the co-product not. Since $\g$ is a graded vector space
it is natural to define the co-product $\Delta$ to be an homorphism
from ${\cal U}(\g)$ to ${\cal U}(\g) \ \hat \otimes \ {\cal U}(\g)$,
where $\hat \otimes$ is the $\mathbb Z_3-$graded tensor product.
This means in particular that for  homogeneous elements
$a_1,a_2,b_1,b_2 \in {\cal U}(\g)$ we have 
$(a_1 \ \hat \otimes \  a_2)(b_1 \ \hat \otimes \  b_2) =
q^{\text{gr}(a_2) \text{gr}(b_1)} a_1 b_1 \ \hat \otimes \ a_2 b_2,$
with $\text{gr}(a)$ the grade of $a$.
We define 

\beqa
\label{coprod}
\Delta 1 = 1 \ \hat \otimes \ 1, \ \Delta g = g \ \hat \otimes \ 1 +
1 \ \hat \otimes \  g, \text{ for any } g \in \g.
\eeqa

\noi
The co-product extends to any elements of ${\cal U}(\g)$ using
\eqref{basis} and $\Delta(g_1 g_2) = \Delta(g_1) \Delta(g_1)$.
In particular this means 
for any $Y_1, Y_2, Y_3 \in \g_1$, from $\Delta(Y_1 Y_2 Y_3)=
\Delta(Y_1) \Delta(Y_2) \Delta(Y_3)=
Y_1 Y_2 Y_3 \ \hat \otimes \ 1 +
Y_1 Y_2 \ \hat \otimes \ Y_3 + q Y_1 Y_3 \ \hat \otimes \ Y_2 
+ q^2 Y_2 Y_3 \ \hat \otimes \  Y_1 +
Y_1 \ \hat \otimes \ Y_2 Y_3 + q Y_2 \ \hat \otimes \ Y_1 Y_3 +
q^2 Y_3 \ \hat \otimes \ Y_1 Y_2 + 1 \ \hat \otimes \ Y_1 Y_2 Y_3,$
that the co-product \eqref{coprod} is compatible
with the Definition \ref{def-flie} {\it i. e. } we have

\beqa
\label{delta-flie}
\{ \Delta Y_1, \Delta Y_2, \Delta Y_3 \}= 
\Delta\{Y_1,Y_2,Y_3\}.
\eeqa

\noi
The antipode is defined by

\beqa
\label{antipod}
S(1)=1, \ S(g)=-g, \text{ for any g } \in \g,
\eeqa

\noi
and the co-unit by

\beqa
\label{counit}
\epsilon(1)=1, \  \epsilon(g)=0, \text{ for any g } \in \g.
\eeqa

\noi
It should be mentioned that a different Hopf algebra  associated to Lie 
algebras of
order three has been introduced  in \cite{ayu},
where the co-product was defined by the usual tensorial product
and the twist (necessary to ensure \eqref{delta-flie})
   was generated by an additional
element.

Now, if for $g \in \g$ we consider the  formal
element  $\text{e}^{g}$ of ${\cal U}(\g)$ (for the moment,
we do not take care of any convergence  problem
and $\text{e}^{g}$ is defined as a formal
series), then, one can show that $\text{e}^{g}$ is a group-like element 
and we have

\beqa\label{grouplike}
\Delta \text{e}^{g} =\text{e}^{g} \otimes \text{e}^{g}, \ S(\text{e}^{g})=\text{e}^{-g}, \ 
\epsilon(\text{e}^{g})=1.
\eeqa

\noi
This means that $e^g$ is a good candidate to define the group associated
to the Lie algebra of order three $\g$. However, before identifying 
$e^g$ to an element of the group associated to $\g$,
the parameters of the transformation should be firstly identified.
It is well known that the parameters associated to a Lie superalgebra
are commuting numbers for the even part and Grassmann numbers for
the odd part. Similarly, we now show that here the parameters associated
to $\g_0$ are commuting numbers although the parameters associated
to $\g_1$ belongs to the $3-$exterior algebra.\\

To identify the parameters of the transformation one needs first to
define ${\cal U}(\g)^*$ the dual of the Hopf algebra ${\cal U}(\g)$.
We define  
the dual basis to be $g^{\vec b, I_\ell}, \vec b \in \mathbb N^{n_0},
\ell \in \mathbb N$ and $I_\ell \in
\left\{1,\cdots,n_1\right\}^\ell \setminus I_{\ell,n_1}$ 
with the natural pairing:

\beqa \label{dual}
\Big<g^{\vec b, I_\ell},g_{\vec a, I_k} \Big>= \delta^{\vec b}{}_{\vec a} \delta^\ell{}_k 
\delta^{I_\ell}{}_{I_k}.
\eeqa

\noi
Denote $\alpha^a$ the dual of $X_{a}$ and  $\theta^i$ the dual of $Y_i$.
The  co-product \eqref{coprod} on ${\cal U}(\g)$ induces the product
on ${\cal U}(\g)^*$.  
 Recall that if $e_i$ constitute
a basis of a Hopf algebra $H$ with co-product $\Delta e_i=\delta_i{}^{jk} e_j \otimes e_k$,
for the dual space $H^\star$ with dual basis $e^i$ we have 
$e^j e^k=\delta_i{}^{jk} e^i$. Using \eqref{coprod} it is easy to see
that the variables $\alpha^a$ commute between themselves and with the
variables $\theta^i$. Furthermore, a little algebra gives that the variables
$\theta$ generate the $3-$exterior algebra and thus, as we now show,
 \eqref{3-ext} is
satisfied for any $\theta^i, \theta^j,\theta^k$. 

If we denote $\theta^1, \theta^{(1,1)}$ the dual variables of
$Y_1$ and $Y_1 Y_1$ the co-product \eqref{coprod} gives

$$\theta^1 \theta^1 =-q^2\theta^{(1,1)}, 
\ \ \theta^1 \theta^{(1,1)}=  \theta^{(1,1)}\theta^1=0, \ \ 
\text{ and thus } \ \ (\theta^1)^3=0.$$

 The case with two different variables, say $\theta^1, \theta^2$ is
more involved. 
We introduce $\theta^1, \theta^2, \theta^{(1,2)}, \theta^{(2,1)},
\theta^{(1,2,1)}, \theta^{(2,1,1)}$ the dual variables of
$Y_1, Y_2, Y_1 Y_2,Y_2 Y_1,
Y_1 Y_2 Y_1, Y_2 Y_1 Y_1$ (since $Y_1 Y_1 Y_2$  is not a Roby element
$- \  1 \le 1 \le 2 \ -$ there is no need to introduce the element
$\theta^{(1,1,2)}$). 
Using 

$$
\Delta (Y_i Y_j) = Y_i Y_j \ \hat \otimes \ 1 +
 Y_i  \ \hat \otimes \ Y_j + q  Y_j  \ \hat \otimes \  Y_i +
 1  \ \hat \otimes \  Y_i Y_j.$$

\noi
 we obtain 

$$
\theta^1 \theta^2 = \theta^{(1,2)} + q \theta^{(2,1)}, \ \
\theta^2 \theta^1 = \theta^{(2,1)} + q \theta^{(1,2)}. \nonumber \\
$$

\noi
Similarly, 

\beqa
\Delta(Y_1 Y_2 Y_1)&=&
Y_1 Y_2 Y_1 \totimes 1 + 
Y_1 Y_2 \totimes Y_1 + q  Y_1 Y_1 \totimes Y_2 + q^2Y_2 Y_1 \totimes Y_1 
\nonumber \\
&+&
Y_1 \totimes Y_2 Y_1 + q Y_2 \totimes Y_1 Y_1 + q^2 Y_1 \totimes Y_1 Y_2 +
1 \totimes Y_1 Y_2 Y_1, \nonumber \\
\Delta(Y_2 Y_1 Y_1)&=& Y_2 Y_1 Y_1 \totimes 1 
-q^2 Y_2 Y_1 \totimes Y_1 + q^2 Y_1 Y_1 \totimes Y_2 \nonumber \\
&+&
Y_2 \totimes Y_1 Y_1 -  Y_1 \totimes Y_1 Y_2 +   1 \totimes Y_2 Y_1 Y_1,
\nonumber 
\eeqa 

\noi
give

\beqa
\theta^1 \theta^1 \theta^2 &=&-\theta^{(1,2,1)}-q\theta^{(2,1,1)}
\nonumber \\
\theta^1 \theta^2 \theta^1&=&2\theta^{(1,2,1)}- \theta^{(2,1,1})
\nonumber \\
\theta^2 \theta^1 \theta^1&=&-\theta^{(1,2,1)} -q^2 \theta^{(2,1,1)}
\nonumber 
\eeqa

\noi
leading to $\theta^1 \theta^1 \theta^2 + \theta^1 \theta^2 \theta^1 +
\theta^2 \theta^1 \theta^1 =0$. 

A similar, but more tedious calculus,
gives along the same lines that for $i \ne j \ne k$ we have
the relation \eqref{3-ext}. Thus, as an algebra we have the following
isomorphism:

$${\cal U}(\g)^* \cong \mathbb C[\alpha^1,\cdots,\alpha^{n_0}]
\otimes \Lambda_3(\mathbb C^{n_1}).$$

\noi
This means in particular that if one wants to associate a group
to the Lie algebra of order 3 $\g=\g_0 \oplus \g_1$ the parameters
of the transformation should be commuting numbers for the elements
of $\g_0$ and should belong to the $3-$exterior algebra for the 
elements of $\g_1$. This means, in relation with \eqref{grouplike}
that one might take something like

\beqa
\label{group}
e^{\alpha^a X_a + \theta^i Y_i},
\eeqa

\noi 
(with $\alpha^a$ commuting numbers and $\theta^i$ the canonical generators
of the $3-$exterior algebra)
to obtain a group. However, here some care as to be
taken.
Firstly  developing $e^{\theta^i Y_1}$  above  
since $\Lambda_3(\mathbb C^{n_1})$
is infinitely generated (see the appendix)
 there   is an infinite number of terms in the series, and we do not
know whether or not the series converge since no convergence properties have been established 
on $\Lambda_3(\mathbb C^{n_1})$, even if $Y_a$ are represented by  
finite-dimensional matrices.
Secondly since there is
 no Baker-Campbell-Hausdorff formul\ae  \ for elements
of $\Lambda_3(\mathbb C^{n_1})$. This means that
 the product of two elements of the type
\eqref{group} is not of the same type. These problems have been solved
in \cite{gr}, but here we propose another way to construct a group
associate to Lie algebras of order 3.

\section{Group associated to Lie algebra of order 3  $-$ the
$\g_{\text{el}}(m_1,m_2,m_3)$ case}

In this section we are constructing explicitly a group
associated to the Lie algebra of order 3 
$\g_{\text{el}}(m_1,m_2,m_3)$
in a straight 
analogy with the construction of the Lie supergroup associated
to the Lie  superalgebra $\mathfrak{gl}(m,n)$.
Using the results of subsection \ref{hopf-sect},
since the parameters of the transformation are
strongly related with the $3-$exterior algebra,  we are considering
matricial groups whose entries belong to the $3-$exterior algebra.

First of all we introduce some set of matrices with coefficients in
$\Lambda_3(\mathbb C^p)$:

\beqa
{\cal M}_k(\Lambda_0) &=&{\cal M}_k(\mathbb C) \ \otimes 
\Lambda(\mathbb C^p)_0,
\nonumber \\
{\cal M}_{k,\ell}(\Lambda_1) &=&{\cal M}_{k,\ell}(\mathbb C) \ \otimes 
\Lambda(\mathbb C^p)_1, \\
{\cal M}_{k,\ell}(\Lambda_2) &=&{\cal M}_{k,\ell}(\mathbb C) \ \otimes 
\Lambda(\mathbb C^p)_2. 
\nonumber 
\eeqa
\bigskip

\noi
A matrix,  {\it e.g.} in ${\cal M}_k(\Lambda_0)$ is given by

\beqa
\label{A0}
A(\Lambda_0)= A_{(0)} + A_{(3)}{}_{ijk} \theta^i \theta^j \theta^k +
A_{(6)}{}_{ijk\ell mn}  \theta^i \theta^j \theta^k 
\theta^\ell \theta^m \theta^n + \cdots
\eeqa

\noi
where the sum above is finite and taken only over Roby elements.

\begin{proposition}
\label{inv}
The matrix given in \eqref{A0} is invertible if and only if
$A_{(0)}$ is invertible.
\end{proposition}

\noi
{\bf Proof:} We introduce 
 $B(\Lambda_0)= B_{(0)} + B_{(3)}{}_{ijk} \theta^i \theta^j \theta^k +
B_{(6)}{}_{ijk\ell mn}\theta^i \theta^j \theta^k  
\theta^\ell \theta^m \theta^n 
+ \cdots$, with the sum taken over Roby elements.
Solving the equation $A(\Lambda_0) 
B(\Lambda_0) = 1$, gives the coefficients of $B(\Lambda_0)$ terms by terms. We thus get for
the two first terms
$B_{(0)}= A_{(0)}^{-1}, B_{(3)}{}_{ijk}= -A_{(0)}^{-1}A_{(3)}{}_{ijk}A_{(0)}^{-1}$.
However, for the higher order terms some care as to be taken. Indeed, the terms of degree six give:

$$(A_{(0)} B_{(6)}{}_{ijk\ell mn} + A_{(6)}{}_{ijk \ell mn}B_0) 
\theta^i \theta^j \theta^k \theta^\ell \theta^m \theta^n +
A_{(3)}{}_{i'j'k'} B_{(3)}{}_{i'' j'' k''} 
\theta^{i'} \theta^{j'} \theta^{k'}  \theta^{i''} \theta^{j''} \theta^{k''}=0.$$ 

\noi But even though $ \left\{\theta^i \theta^j \theta^k \theta^\ell \theta^m \theta^n\right\} $ are by definition of
$A(\Lambda_0)$ and $B(\Lambda_0)$ independent elements of $\Lambda_3(\mathbb C^p)$, in general it is not  true for 
$\left\{\theta^{i'} \theta^{j'} \theta^{k'}  \theta^{i''} \theta^{j''} \theta^{k''}\right\}$. Indeed, 
if the sequences  $(i',j',k'), (i'',j'',k'') \in
\left\{1,\cdots,p \right\}^3 \setminus I_{3,p}$
this does not mean that the sequence  $(i',j',k',i'',j'',k'') \in  
\left\{1,\cdots, p \right\}^6 \setminus I_{6,p}$,
as can be seen with $(1,2,1)$ and $(1,2,1)$. 
Decomposing this last terms in 
the basis of $\Lambda_3(\mathbb C^p)$ we have 
$$ \theta^{i'} \theta^{j'} \theta^{k'}  \theta^{i''} \theta^{j''} \theta^{k''}
 =
C^{i' j' k' i'' j'' k''}{}_{ijk\ell mn} \theta^i \theta^j \theta^k \theta^\ell \theta^m \theta^n$$ 

\noi
(where the sum is taken over $(ijk\ell mn) \in 
\left\{1,\cdots, p \right\}^6 \setminus I_{6,3}$)
and we get

$$B_{(6)}{}_{ijk\ell mn}=-A_{0}^{-1} A_{(6)}{}_{ijk \ell mn} A_{(0)}^{-1} - 
C^{i' j' k' i'' j'' k''}{}_{ijk\ell mn}
A_{(0)}^{-1} A_{(3)}{}_{i' j' k'} A_{(0)}^{-1} A_{(3)}{}_{i'' j'' k''} A_0^{-1}.$$

\noi
This procedure extends simply order by order.
 QED. \\

We denote by $GL(m_1,\Lambda_0)$ the set of invertible matrices
of ${\cal M}_{m_1}(\mathbb C,\Lambda_0)$.

\begin{proposition}
Let  $A_0 \in GL(m_1,\Lambda_0),
A_1 \in GL(m_2,\Lambda_0),
A_2 \in GL(m_3,\Lambda_0),$
$B_0 \in  {\cal M}{}_{m_3,m_1}( \Lambda_1),
B_1 \in {\cal M}{}_{m_1,m_2}(\Lambda_1),
B_2 \in  {\cal M}{}_{m_2,m_3}(\Lambda_1)$
 and 
$C_0 \in  {\cal M}{}_{m_2,m_1}(\Lambda_2),
C_1 \in  {\cal M}{}_{m_3,m_2}(\Lambda_2),
C_2 \in  {\cal M}{}_{m_1,m_3}(\Lambda_2)$, then
the $m_1 \times m_2 \times m_3$  matrix with coefficients
in $\Lambda_3(\mathbb C^p)$ 

\beqa\label{group1}
M=\begin{pmatrix} A_0&B_1&C_2\\
                  C_0&A_1&B_2\\
                  B_0&C_1&A_2
\end{pmatrix},\eeqa

\noi
is  invertible.

\end{proposition}

{\bf Proof: }
By definition, it is obvious to see that if $M, N $
are two matrices of the type above  
then the product 
$MN $ is also as above.
We now show that the matrix $M$ is invertible. Define

$$N=\begin{pmatrix} A'_0&B'_1&C'_2\\
                  C'_0&A'_1&B'_2\\
                  B'_0&C'_1&A'_2
\end{pmatrix},$$
and solving MN= 1, we get

\beqa\label{i3}
A_0'&=&\Big\{A_0-C_2(A_2-C_1A_1^{-1} B_2)^{-1}(B_0-C_1A_1^{-1}C_0)  \nonumber \\
&& \hskip .75truecm -B_1A_1^{-1}\left[C_0 -B_2(A_2-C_1 A_1^{-1}B_2)^{-1}
(B_0-C_1A_1^{-1}C_0)\right]
\Big\}^{-1}, \nonumber \\
B_0'&=&-(A_2-C_1A_1^{-1} B_2)^{-1}(B_0-C_1A_1^{-1}C_0)A_0', \nonumber \\
C_0'&=&-A_1^{-1}\Big[C_0 -B_2(A_2-C_1 A_1^{-1}B_2)^{-1}(B_0-C_1A_1^{-1}C_0)\Big
]A_0', 
\nonumber \\
A_1'&=&\Big\{A_1-C_0(A_0-C_2A_2^{-1} B_0)^{-1}(B_1-C_2A_2^{-1}C_1) \nonumber \\
&& \hskip .75truecm -B_2A_2^{-1}\left[C_1 -B_0(A_0-C_2 A_2^{-1}B_0)^{-1}
(B_1-C_2A_2^{-1}C_1)\right]
\Big\}^{-1}, \nonumber \\
B_1'&=&-(A_0-C_2A_2^{-1} B_0)^{-1}(B_1-C_2A_2^{-1}C_1)A_1',  \\
C_1'&=&-A_2^{-1}\Big[C_1 -B_0(A_0-C_2 A_2^{-1}B_0)^{-1}(B_1-C_2A_2^{-1}C_1)\Big
]A_1',
 \nonumber \\
A_2'&=&\Big\{A_2-C_1(A_1-C_0A_0^{-1} B_1)^{-1}(B_2-C_0A_0^{-1}C_2) \nonumber \\
&& \hskip .75truecm -B_0A_0^{-1}\left[C_2 -B_1(A_1-C_0 A_0^{-1}B_1)^{-1}
(B_2-C_0A_0^{-1}C_2)\right]
\Big\}^{-1}, \nonumber \\
B_2'&=&-(A_1-C_0A_0^{-1} B_1)^{-1}(B_2-C_0A_0^{-1}C_2)A_2', \nonumber \\
C_2'&=&-A_0^{-1}\Big[C_2 -B_1(A_1-C_0 A_0^{-1}B_1)^{-1}(B_2-C_0A_0^{-1}C_2)\Big
]A_2'. \nonumber 
\eeqa

\noi
The formul\ae   \ above make sense because of Proposition  \ref{inv}.
QED

However, it can happen that the sum defining the inverse matrix
of a given matrix $A(\Lambda_0)$ 
is  infinite even if
$A(\Lambda_0)$ is defined by a finite sum. Indeed, we have 
$(A_0 + A_1  \theta^3 \theta^2 \theta^1)^{-1} 
 = \sum_{k \ge 0} (-A_0^{-1} A_1)^k A_0^{-1} 
( \theta^3 \theta^2 \theta^1)^k$, we thus restrict ourselves to the set of matrices such that
$A(\Lambda_0)$ and its inverse $A^{-1}(\Lambda_0)$ are given by finite sums.
Looking to the sum above, $A(\Lambda_0)^{-1}$ becomes a finite sum if there is some nilpotent elements.
For instance this  happens if $A_0^{-1} A_1$ is nilpotent. It  can also happen that some power
of elements in $\Lambda_3(\mathbb C^p)$ vanishes. For instance, since
$(\theta^2 \theta^2 \theta^1)^2=-\theta^2 \theta^2(\theta^2\theta^1\theta^2 
+ \theta^2 \theta^2 \theta^1)\theta^1 =0$,
we have  $(A_0 + A_1  \theta^2\theta^2 \theta^1)^{-1} 
 =A_0^{-1}  -A_0^{-1} A_1 A_0^{-1} 
 \theta^2 \theta^2 \theta^1$. 
Thus in order that the formul\ae \ above make sense we consider the 
set of matrix  \eqref{group1} such that the various matrices together with their inverse
(see  \eqref{i3})  are given by finite sum. 
\\

\begin{definition}

The set of matrices of the forme

\beqa\label{group2}
M=\begin{pmatrix} A_0&B_1&C_2\\
                  C_0&A_1&B_2\\
                  B_0&C_1&A_2
\end{pmatrix},\eeqa

\noi
where the matrices $A_0 \in GL(m_1,\Lambda_0),
A_1 \in GL(m_2,\Lambda_0),
A_2 \in GL(m_3,\Lambda_0),$
$B_0 \in  {\cal M}{}_{m_3,m_1}( \Lambda_1),
B_1 \in {\cal M}{}_{m_1,m_2}(\Lambda_1),
B_2 \in  {\cal M}{}_{m_2,m_3}(\Lambda_1),$ 
$C_0 \in  {\cal M}{}_{m_2,m_1}(\Lambda_2),
C_1 \in  {\cal M}{}_{m_3,m_2}(\Lambda_2),
C_2 \in  {\cal M}{}_{m_1,m_3}(\Lambda_2)$,
together with the matrices
 $A'_0 \in GL(m_1,\Lambda_0),
A'_1 \in GL(m_2,\Lambda_0),
A'_2 \in GL(m_3,\Lambda_0),$
$B'_0 \in  {\cal M}{}_{m_3,m_1}( \Lambda_1),
B'_1 \in {\cal M}{}_{m_1,m_2}(\Lambda_1),
B'_2 \in  {\cal M}{}_{m_2,m_3}(\Lambda_1),$
$C'_0 \in  {\cal M}{}_{m_2,m_1}(\Lambda_2),
C'_1 \in  {\cal M}{}_{m_3,m_2}(\Lambda_2),
C'_2 \in  {\cal M}{}_{m_1,m_3}(\Lambda_2)$,
(see \eqref{i3}) are given by finite sums  is a group
we denote $GL_f(m_1,m_2,m_3,\Lambda)$.
\end{definition}

We now show, in relation with the previous
section,  that this group in non-empty. 
Indeed, if we take 

$$g=e^{A} \prod_k e^{B_k},$$

\noi
 with
$A$ an arbitrary matrix of $\mathfrak{gl}(m_1) \times \mathfrak{gl}(m_2) \times
\mathfrak{gl}(m_3)$ and $B_k$ an arbitrary nilpotent matrix of $\mathfrak{gl}(m_1,m_2,m_3)_1
\otimes \Lambda_3(\mathbb C^p)_1$ then $g\in GL_f(m_1,m_2,m_3,\Lambda)$. There is several type of matrix
$B_k$. For instance one can take $B_k= Y P(\theta)$ with (i) $Y$ a nilpotent
matrix of $\mathfrak{gl}(m_1,m_2,m_3)_1$  and $P$ an arbitrary
polynomial of degree $1 \text{ mod } 3$ in $\theta$ or (ii) $Y$ an arbitrary element
of  $\mathfrak{gl}(m_1,m_2,m_3)_1$ and $P$
a nilpotent
polynomial of degree $1 \text{ mod } 3$ in $\theta$ (some power of $P$ vanishes).
More details will be given in \cite{gr}.
\\

To close this section,
we now  show that the  algebra corresponding to the
group $GL_f(m_1,m_2,m_3,\Lambda)$ is $\mathfrak{gl}_{\text{el}}(m_1,m_2,m_2)$.
Indeed, if we take the infinitesimal version of $GL_f(m_1,m_2,m_3,\Lambda)$ 
{\it i.e.} we assume that
the parameters of the transformation characterised by $\theta$ goes to zero 
we have

\beqa
\label{group-alg}
\begin{array}{llll}
A_0&=A_0{}_{(0)} + A_0{}_{(3)}{}_{ijk} \theta^i \theta^j \theta^k  + \cdots 
&\to&A_0{}_{(0)} + {\cal O}((\theta)^2) \\
B_0&=B_0{}_{(1)}{}_i\theta^i + A_0{}_{(4)}{}_{ijk\ell} \theta^i \theta^j \theta^k
\theta^\ell + \cdots 
&\to&A_0{}_{(1)}{}_i \theta^i + {\cal O}((\theta)^2) \\
C_0&=C_0{}_{(2)}{}_{ij}\theta^i \theta^j + C_0{}_{(5)}{}_{ijk \ell m} 
\theta^i \theta^j \theta^k
\theta^\ell \theta^m + \cdots 
&\to&0 + {\cal O}((\theta)^2)\\
\end{array}
\eeqa

\noi
 with $A_0{}_{(0)} \in GL(m_1)$ (and similar relations  for 
$A_1,A_2, B_1,B_2,C_1,C_2$). 
Since $\mathfrak{gl}(m_1)$ is the 
Lie algebra of $GL(m_1)$, the relations \eqref{group-alg} ensure
that the algebra associated to $GL_f(m_1,m_2,m_2,\L ambda)$ is the
Lie algebra of order 3 $\mathfrak{gl}_{\text{el}}(m_1,m_2,m_3)$.
In fact the Hopf algebra technics of the previous sub-section
gives the converse correspondence {\it i.e.} enables us to
establish that the group associated to $\mathfrak{gl}_{\text{el}}(m_1,m_2,m_3)$
is $GL_f(m_1,m_2,m_3,\Lambda)$ \cite{gr}.

{\bf  Acknowledgments}
J. Lukierski is gratefully acknowledge for discussions.

\appendix
\section{The three-exterior algebra}
The three-exterior  algebra has been introduced by Roby \cite{roby}
in 1970 as a possible (cubic) generalisation of the Grasmann algebra.
The three-exterior algebra $\Lambda_3(\mathbb K^n)$ 
is the unitary (with unit denoted by $1$)
$\mathbb K-$algebra ($\mathbb K=\mathbb R$ or 
$\mathbb K = \mathbb C$) generated by $n$ canonical generators
$\theta^1,\cdots, \theta^n$ submitted to the relation

\beqa
\label{3-ext}
 \theta^i \theta^j \theta^k + 
\theta^j \theta^k \theta^i + 
\theta^k \theta^i \theta^j +
  \theta^i \theta^k \theta^j + 
\theta^j \theta^i \theta^k + 
\theta^k \theta^j \theta^i=0.
\eeqa

\noi
This algebra which can be seen as a possible 
generalisation of the Grassmann algebra is indeed very different. Because
the algebra
 $\Lambda_3(\mathbb K^n)$ is defined through cubic  
relations the number of independent monomials
increases with polynomial's degree 
(for instance, $(\theta^1)^2 \theta^2$ and $\theta^1 \theta^2 \theta^1$ are 
independent). 
This means that we do not have  enough constraints among the 
generators to order them in some fixed way and, as a consequence,
$\Lambda(\mathbb K^n)$ turns out to be an infinite dimensional algebra.
To characterise precisely $\Lambda(\mathbb K^n)$ it should be 
interesting to obtain a basis. This is complicated by the fact that we
have some cubic relations among the generators and for instance
the three element $(\theta^1)^2 \theta^2, \theta^1 \theta^2 \theta^1$ and
$\theta^2 (\theta^1)^2$ are not independent since there sum is zero.
To characterise the set of independent elements one needs a
definition.\\

\noi
{\bf Definition A1}
{\it 
The sequence 
 $(i_1, i_2,\cdots i_k) \in \left\{1,\cdots, n \right\}^k , \ k \ge 3$
 has a 
rise of length $3$
if there exists $0 \le \ell \le k-3$ such that $i_{\ell+1} \le i_{\ell+2}
\le i_{\ell+3}$. We denote by $I_{k,n}$ the set of
$k$ indices which has a rise of length $3$.
}
\medskip

\noi
{\bf Theorem A2} {\it (N. Roby \cite{roby})
The elements

$$ \left\{ \theta^{i_1} \theta^{i_2} \cdots \theta^{i_k},
k \in \mathbb N, (i_1,\cdots, i_k) \in 
\left\{1, \cdots n \right\}^k \setminus I_{k,n} 
\right\},$$

\noi
constitute a basis of $\Lambda_3(\mathbb K^n)$.}

\medskip 
For instance if $n=2$ a basis of $ \Lambda_3(\mathbb C^2)$ is given by 

\beqa
&&1, \nonumber \\ 
&&\theta^i, \ \ i=1,2,  \nonumber \\  
&&\theta^i \theta^j, \ \ i,j=1,2, \nonumber \\
&& \theta^i(\theta^2 \theta ^1)^n, \ \ (\theta^2 \theta ^1)^n\theta ^i,\ \ 
 i=1,2, \ \  n \ge 1, 
\nonumber \\
&& \theta^i(\theta^2 \theta^1)^{n-1}\theta^j, \ \  i,j=1,2, \ \
(\theta^2 \theta^1)^n, \ \ n \ge 2.
\nonumber
\eeqa

\noi
From now on we call  Roby elements the elements
$\theta^{I_k} = \theta^{i_1} \cdots \theta^{i_k}$ such that 
$I_k \in \left\{1,\cdots,n\right\}^k \setminus
I_{k,n}$.

The $3-$exterior algebra is a $\mathbb Z_3-$graded algebra. Indeed, if
we define $\Lambda(\mathbb K^n)_i, i=0,1,2$ the set of elements
of degree $i$ mod $3$ we have 
$\Lambda_3(\mathbb K^n)_i \Lambda_3(\mathbb K^n)_j \subseteq 
  \Lambda_3(\mathbb K^n)_{i+j}$ and
$  \Lambda_3(\mathbb K^n)= \Lambda_3(\mathbb K^n)_0 \oplus
 \Lambda_3(\mathbb K^n)_1 
\oplus \Lambda_3(\mathbb K^n)_2$.

Finally let us mention that 
the results given in this appendix extend to any $F>2$ \cite{roby}.


\bigskip\bigskip
\begin{thebibliography}{19}
\bibitem{bg1}
Bars I  and  Gunaydin M 1979 
  {\it  J.\ Math.\ Phys.\ }  {\bf 20} 1977-1993.
\bibitem{bg2}
Bars I  and  Gunaydin M 1980 
{\it   Phys.\ Rev.\  D } {\bf 22}  1403-1413.
\bibitem{k1}
Vainerman L and Kerner R 1995 {\it J. Math. Phys}
{\bf 37} 2553-2565.
\bibitem{k2}
Kerner R 1997 {\it Class. and Quantum Grav.} {\bf 14 (1A)} A203-A225.
\bibitem{k3}
 Borowiec A,  Bazunova  N and 
Kerner R 2004 {\it  Lett. Math. Phys.}
{\bf  67}  195-206.
%
\bibitem{r}
Rausch de Traubenberg M,
 {\it Clifford algebras, supersymmetry and 
$\mathbb Z_n-$symmetries: Applications in
field theory,}
  arXiv:hep-th/9802141 (Habilitation Thesis).
\bibitem{f}
Filipov V. T.  1985 {\it Sibirsk. Math. Zh} {\bf 26} 126-140.
%
\bibitem{g}
Gnedbaye  A. V.  1995 {\it C. R. Acad. Sci.  Paris S\'er. I Math.}
{\bf   321}   147-152.
%
\bibitem{mv}
Michor  P. W. and  Vinogradov  A. M. 1996 
{\it Rend. Sem. Mat. Univ. Politec. Torino} {\bf   54} 373-392.
%
\bibitem{rs1}
 Rausch de Traubenberg M and Slupinski M J 2000
{\it J. Math. Phys. } {\bf 41} 4556-4579,
[arXiv:hep-th/9904126].
\bibitem{rs2}
 Rausch de Traubenberg M and Slupinski M J  2002 
{\it J. Math. Phys. } {\bf 43} 5145-5160,
[arXiv:hep-th/0205113].
\bibitem{cubic1}
Mohammedi N,   Moultaka  G and  Rausch de Traubenberg M  2004 
{ \it Int. J. Mod. Phys. }
{\bf A19}  5585-5608, [arXiv:hep-th/0305172].
\bibitem{cubic2}
Moultaka G,  Rausch de Traubenberg M  and Tanasa A (2005)
{\it Int. J. Mod. Phys. } {\bf A20}  5779-5806, 
[arXiv:hep-th/0411198]. 
\bibitem{pform}
Moultaka G, Rausch de Traubenberg M and Tanasa A 2004
 { Proceedings of the XIth International Conference Symmetry Methods 
in Physics, Prague 21-24 June 2004}, [arXiv:hep-th/0407168].
\bibitem{grt1}
Rausch de Traubenberg M 2006 {\it to appear in the Proocedings of the
International Conference on Symmetry Methods in Physics (SYMPHYS-12), Yerevan, Armenia, 3-8 Jul 2006},
 [arXiv:hep-th/0612204].
\bibitem{grt2}
Goze M, Rausch de Traubenberg M and Tanasa A 2007 {\it J. Math. Phys.}
{\bf 48} 093507, [arXiv:math-ph/0603008].
%
\bibitem{roby} Roby N 1970 {\it Bull. Sc. Math.} {\bf 94} 49-57.
%
\bibitem{gr}
Goze M and Rausch de Traubenberg M, {\it in preparation}.
%
\bibitem{ayu}
Ahmedov H, Yildiz A  and   Ucan Y 2001  
{ \it J.\ Phys.}  { \bf A 34} 6413-6423,
[arXiv:math.RT/0012058].
\end{thebibliography}
\end{document}